\documentclass[twocolumn,showpacs,preprintnumbers,floatfix,prl]{revtex4}

\usepackage{graphicx}
\begin{document}

\title{Antibubbles : evidences of a critical pressure}

\author{S.Dorbolo and N.Vandewalle}

\affiliation{GRASP, Institut de Physique B5, Universit\'e de Li\`ege, \\ B-4000 Li\`ege, Belgium.}

\begin{abstract}
We present experimental investigations of antibubbles.  Such an unusual fluid object is a thin spherical air shell surrounding a liquid globule.  We explain how to produce them and we study their stability.  By overweighting antibubbles with a small amount of salt, they sink and pop at a definite depth.  A critical depth related to a critical pressure has been found.  A modified Laplace law describes the air shell thickness evolution with respect to pressure.  This law combined with surfactant layers interaction allows to explain the critical depth for antibubble stability.

 \end{abstract}

      \maketitle         

A bubble is a thin film of liquid enclosing a gas pocket. The physics of bubbles is well established, and a large number of text books concerns this subject \cite{weaire}.  However, it is not generally known that one can also have a thin film of gas enclosing a liquid (see Fig.1). This unusual object was first reported by Hughes and Hughes \cite{hughes}, and was thereafter coined {\it antibubble} by Connett \cite{stong}.  Since these reports, antibubbles were mainly a curiosity but questions remain about their stability and their physics.

\begin{figure} \begin{center} 

\includegraphics[width=6cm]{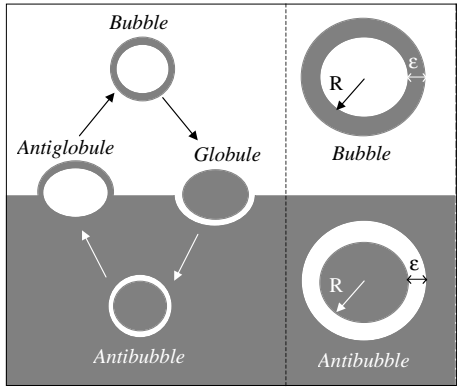}
\caption{(right) Schematic representation of the different configurations of a liquid in contact with a gas; the arrows indicate the possible transformations. (left) Comparison of a bubble and an antibubble characterized by a inner radius $R$; $\varepsilon$  being  the thickness of the minority fluid.} \end{center} \end{figure}

Figure 1 summarizes the different configurations of spherical fluid interfaces.  If a single interface is considered, we have either a drop of water (so-called globule) or a sphere of gas in a liquid (so-called antiglobule).  Moreover the well known bubble constituted of a thin spherical layer of liquid separating the gas at different pressures,  has its symmetric object called antibubble.  Note that the bubble and the antibubble are composed of two interfaces.  The arrows in Fig.1 represent the different possible transformations between the different fluid entities.  For example, an antiglobule may be extracted from the liquid to become a bubble and a popping bubble becomes a globule (a small drop of liquid).  The symmetric transformation from a globule into an antibubble and finally into an antiglobule will be explained below.

\begin{figure} \begin{center} 

\includegraphics[width=6cm]{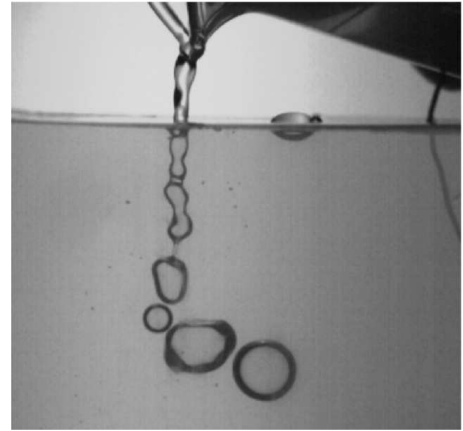}
\caption{Formation of an antibubble from a b\'echer.  The incoming liquid streamer breaks up into several antibubbles.  At the liquid surface, a liquid globule floats.  The metallic connection between the incoming liquid and the liquid of the vessel can be seen on the right.} \end{center} \end{figure}

A plexiglass vessel has been specially built for our purpose. This experimental cell is 200 cm tall and has a base of 15 x 10 cm$^2$. The cell is open on the top and is completely filled with a liquid mixture (30 l water and 245 ml dishwashing detergent).  The cell is placed in a larger tank for preventing leaks and a continuous flow is maintained with a pump in order to keep clean the liquid surface at the top of the cell. With the help of a b\'echer, we drop globules of liquid on the surface, and by adjusting the liquid jet the antibubbles are formed when the globules sink.  Figure 2 presents a streamer of liquid that feeds forming antibubbles below the surface.  The antibubble production zone extends from zero to 5 cm below the surface. In order to prevent any electrical potential difference accross the thin air film, a copper wire connects the liquid in the b\'echer and the liquid in the experimental cell.  We have observed that the air film is more stable when this wire is present.  This point is important and will be discussed below.  Antibubbles have been produced with diameters up to 1.6 cm. It was impossible to create larger antibubbles with our setup and physical reasons for that upper limit are given below.

The mechanism of antibubble production can be decomposed as follows.  A globule is first created at the surface of the liquid.  The incoming liquid flow which feeds the globule has a certain velocity and the globule sinks.  The flow forms a streamer inside the liquid.  Rayleigh-Plateau instabilities appears along this streamer (see Fig. 2) \cite{laplace}.  The streamer breaks up and forms antibubbles. 

\begin{figure} \begin{center} 
\includegraphics[width=6cm]{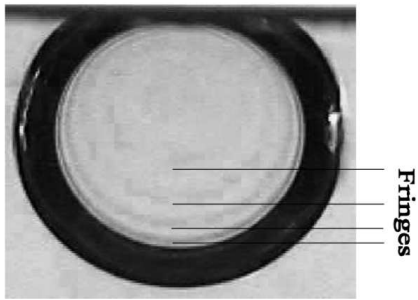}
\caption{Antibubble located just below the liquid surface.  The diameter of the antibubble is 1.5 cm.  Interference fringes similar to Newton's rings are emphasized.  The dark circumference is due to some total reflections.} \end{center} \end{figure}

Once formed, the antibubbles rise in the liquid.  Indeed, they are dragged towards the surface by buoyant forces.  Since the air volume contained in an antibubble is quite small, the motion is slow and the antibubbles keep a spherical shape. When reaching the surface, the antibubbles touch the air/liquid interface where they stay for a long time (a few minutes) before naturally popping due to some aging.  Such an antibubble `glued' just below the liquid surface is shown in Fig.3.  By popping antibubbles with a needle and by comparing the diameters of the antibubble and the resulting air spherical pocket (antiglobule), we estimated the average thickness of the air film to be $\varepsilon = 3 \mu$m.  This distance is quite small but similar to the film thickness of bubbles.  In the case of bubbles, the liquid film leads to interferences and subsequent horizontal colored fringes appear.  Colored interference fringes can be seen in antibubbles as well (Fig.3).  Their shape similar to that found in Newton rings problem can be observed \cite{newton}.  The thick black circumference is due to total reflections of light on the first diopter liquid-air shell.

\begin{figure} \begin{center} 
\includegraphics[width=8cm]{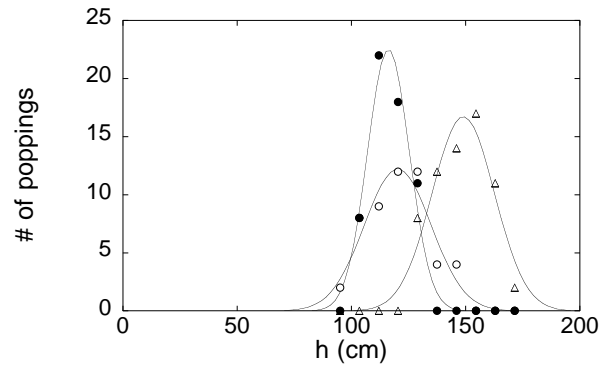}
\caption{Distribution of the critical depths for three experimental sessions (circles, triangles and squares).  Solid curves represent gaussian fits of the data.} \end{center} \end{figure}

In order to avoid floating antibubbles, we added a small amount of salt in some antibubbles.  Such `heavy antibubbles' reach rapidly a constant free falling velocity. The largest antibubbles sink with a velocity around 10 cm/s while the smallest ones (diameter 0.5 cm) fall at 2 cm/s.  In the first series of experiments, we surprisingly observed that the great majority of the antibubbles breaks around a critical depth $h$=120 cm wathever their size and whatever their falling velocity.  Other experiments have been performed the next days.  Figure 4 presents three experiments performed at different days (circles, triangles and squares); the preparation of the liquid and the antibubble production process had been the same.  About 60 antibubbles of different sizes have been sunk during each serie.  The distribution of the critical depths has been reported in Fig.4 and fitted by gaussian laws (solid lines).  From one day to another, the mean critical depth $\langle h \rangle$ changes, namely 120 cm, 116 cm, and 148 cm for the circles, triangles and squares respectively.  Those variations have been attributed to the variations of the atmospheric pressure.  Let us remind that a typical daily variation of the atmospheric pressure (about 20 hPa) corresponds to a depth of 20 cm in water.  {\it Thus, the stability of an antibubble seem to depend on the external pressure}. 

\begin{figure} \begin{center} 
\includegraphics[width=8cm]{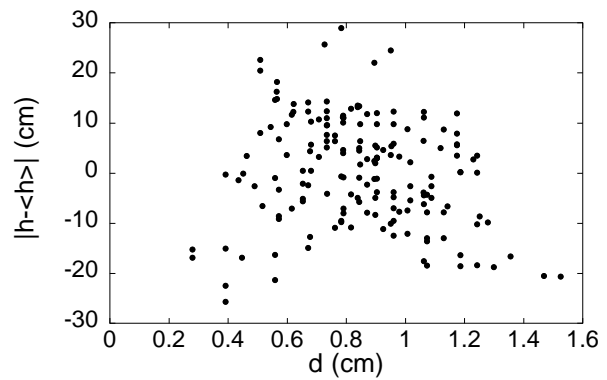}
\caption{Relative critical depths $h-<h>$ as a function of their diameter $d$.} \end{center} \end{figure}

The limit speed of a free falling object in a fluid is mainly determined by its diameter $d$ and to the viscosity.  One can wonder whether a relation exists between the antibubble diameter and its critical depth.  In order to better compare all series of data, the relative critical depths defined as $h-\langle h \rangle$ are plotted versus the sizes of the antibbuble in Fig.5.  The distribution of the results is isotropic around the mean diameter of 0.85 cm.  Furthermore, the distribution of relative critical depths is plotted for the large ($>$ 0.85 cm) and the small ($<$ 0.85 cm) antibubbles in Fig.6.  Gaussian distributions are centered around $0$.  {\it Thus, the stability of an antibubble seems to be size independent}.

\begin{figure} \begin{center} 
\includegraphics[width=8cm]{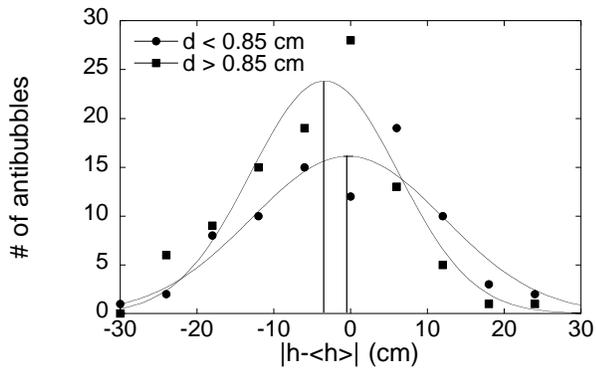}
\caption{Distributions of relative critical depths reported for big ($d > 0.85$ cm) and small ($d < 0.85$ cm) antibubbles.} \end{center} \end{figure}

An antibubble lift has been also built (Fig.7).  This system is directly inspired from the diver bell.  The antibubble is trapped under the bell and could be placed at some definite depth.  The advantage is twofold since no additional salt is needed and the speed is controlled whatever their size.  The critical depth has been found to be the same as before.  {\it Thus, the stability of an antibubble seems to be independent of the free falling velocity}.

The well-known Laplace law applies for bubbles and links the radius $R$ of the bubble to the difference of air pressure $\Delta p$ across the thin film of liquid.  Let us consider a surface element $dA$ of the soap film.  Two forces apply on this surface.  The first one is due to the difference of pressure across the film and the second one comes from the surface tension $\gamma$.  As the bubble grow, both surface and volume variations have to be taken into account.  One has a free energy variation $dF = -(\Delta p) dV + 2 \gamma dA$. The factor 2 in the surface term comes from both gas-liquid and liquid-gas interfaces. Bubble equilibrium reads $dF=0$. When both volume and surface variations are expressed as a function of $R$ and $dR$, the equilibrium condition reads
\begin{equation}
R = \frac{4 \gamma}{ \Delta p}
\end{equation} that one can find in every physics textbook \cite{laplace}.  The Laplace equation is important because it contains the main ingredients which control the physics of a wide range of liquid systems: from droplets to emulsions and foams \cite{weaire}. 

\begin{figure} \begin{center} 
\includegraphics[width=6cm]{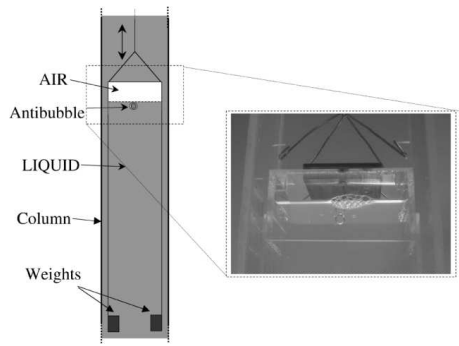}
\caption{Sketch and picture of the antibubble lift.  The antibubble are trapped under the air bell.  The pressure is increased with an arbitrary speed by placing the lift at a definite depth in the column.} \end{center} \end{figure}

For an antibubble, the situation is quite different because the compressibility of the central globule is negligeable with respect to the compressibility of the spherical air shell. That means that the Laplace law has to be modified for antibubbles.  Let $R$ being the radius of the 
liquid globule and $\varepsilon$ being the thickness of the spherical air 
shell. The Laplace equations for both interfaces, i.e. starting from the center liquid-air and then air-liquid, are $\Delta p_1  =  2 \gamma/R$ and $\Delta p_2  =   2 \gamma/(R+\varepsilon)$ respectively.  If one considers any external pressure variation, $\varepsilon$ becomes the relevant parameter.  Indeed, the only way to equilibrate both work and surface energy in an antibubble is to expand or reduce the thickness $\varepsilon$ of the air film while the radius $R$ of the central globule remains constant.

Only the possible pressure difference $\Delta p_2$ accross the external air/liquid interface contributes to free energy variations. In that case, antibubble equilibrium leading to a modified Laplace law 
\begin{equation} \label{mll}
\varepsilon= \frac{2 \gamma}{\Delta p_2} - R
\end{equation} The origin of this pressure is the curvature of the equivalent antiglobule of radius $R+\varepsilon$. 

Assuming that the ideal gas law applies, $p V$ is a constant.  The internal pressure $p$ is given by the sum of three terms : (i) the atmospheric pressure $p_0$, (ii) the hydrostatic pressure $\rho g h$ and (iii) the pressure $\Delta p$ due to the surfactant membrane given by Eq.(\ref{mll}).  The relation between the pressure in the air shell and $\varepsilon$ is thus given by
\begin{equation}\label{mll2}
\left (p_0+\rho g h+\frac{2 \gamma}{R}\right) \varepsilon = c
\end{equation} since the volume of the air shell is $4 \pi R^2 \varepsilon$.  The parameter $c$ is a constant.  In a scope of comparison, the equation relative to bubbles reads
\begin{equation}
\left(p_0+\frac{4 \gamma}{R}\right ) R^3 = c
\end{equation}  where the relevant variable is the bubble radius $R$.

Using Eq.(\ref{mll2}), the variation of $\varepsilon$ with depth can be estimated.  A $10\%$ change of $\varepsilon$ is found when the antibubble sinks to a depth of 150 cm in the column.  This variation is enough to observe a displacement of interference fringes  during the free falling and is enough to cause the popping at this depth.  This calculation gives an order of magnitude of the thickness variation of the air shell.  Nevertheless, Eq.(\ref{mll}) surestimates $\varepsilon$.  Indeed, an important effect has to be taken into account as far as two layers are in interaction : {\it the disjoining pressure} $p_{dis}$.  This term should be added to the sum of the different pressure in the Eq.(\ref{mll2}).

In Fig.8, a soap film is represented on the left.  The surfactant molecules are composed of a hydrophilic head and a hydrophobic tail.  To form a film, the surfactant molecules take place along the liquid film with the head towards the liquid.  During the film thinning, the electrostatic repulsion increases between hydrophobic heads.  This force is counterbalanced by Van der Waals attractive forces.  This creates the so-called disjoining pressure that can be positive or negative according to the thickness of the film \cite{weaire,nato}.  The relevant parameters are the thickness of the film, the nature of the hydrophobic heads (here sodium ion) and the medium between the layers.

On the other hand, in Fig. 8 (right), an air film is represented.  In this case, the tails of the surfactant molecules are in opposition along the air shell.  That means that the interaction forces are different from that of a soap film.  For an air film, the Van der Waals interaction between the tails of surfactant molecules becomes more important in front of the electrostatic contribution of the heads.  Moreover, the medium enclosed between the layers is air.  The dielectric constant $\chi$ of the air is equal to 1, at least 80 times lower than in the soap liquid.  No electrostatical screening effects are expected in the air shell since no electrolytes are present there.  This explains the high sensitivity of antibubbles to any difference of potential between the liquid contained in the globule and the surrounding liquid.  

The stability of the antibubble is given by the balance between Van der Waals attractive forces (in $1/\varepsilon^3$ \cite{nato}) and electrostatic repulsion forces (in $1/(R+\varepsilon)^2$).  The interaction between both surfactant layers is attractive below a certain value of the interlayer distance, i.e. a critical thickness $\varepsilon_c$.  Below $\varepsilon_c$, the equilibrium cannot be maintained anymore and the air film collapses.  The popping mechanism is the following : an increase of the external pressure decreases the thickness of the air shell, this implies an increase of the Van der Waals interactions between the layers until this force becomes so high that the air film breaks up.

\begin{figure} \begin{center} 
\includegraphics[width=7cm]{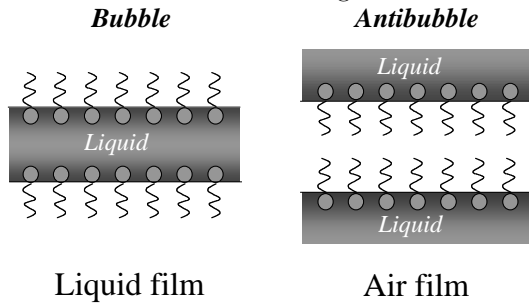}
\caption{Molecular structure of a soap film as found in bubble (left) and of an air film from an antibubble (right).} \end{center} \end{figure}

In summary, our experimental results support the modified Laplace law that governs the air shell thickness.  The popping of antibubbles is mainly controlled by $\varepsilon$ and molecular interactions along the air film.  As the pressure difference increases, $\varepsilon$ decreases to a certain critical value.  This value correspond to the smallest possible distance between two surfactant layers.  Below this value, Van der Waals forces dominate and the air film collapses.  Our study is the basis for further experimental works on antibubbles, antifoams, and other exotic fluid entities.

SD would like to thank FNRS for financial support.  This work has been also supported by the contract ARC 02/07-293.

\end{document}